\documentclass[referee,useAMS,usenatbib]{mn2e} 
\usepackage{graphicx}

\title{Analytical studies of particle dynamics in bending waves in planetary rings}
\author[Abhijit Bhattacharyya and Sandip K Chakrabarti]
{Abhijit Bhattacharyya$^1$\thanks{E-mail: b\_abhijit2k@yahoo.co.in} 
and Sandip K.\ Chakrabarti$^{2,1}$\thanks{E-mail: chakraba@bose.res.in} \\
$^{1}$Centre for Space Physics, Chalantika 43, Garia Station Road, 700 084, India\\
$^{2}$ S. N. Bose National Center for Basic Sciences, JD-III, Salt Lake, Kolkata, 700 098, India\\
}

\pagerange{\pageref{firstpage}--\pageref{lastpage}} \pubyear{2004}

\begin{document}
\label{firstpage}
\maketitle

\begin{abstract}
Particles inside a planetary ring are subject to forcing due to the central planet, moons in 
inclined orbits, self-gravity of the ring and other forces due to radiation drag, collisional effects and 
Lorentz force due to magnetic field of the planet. We write down the equations of motion of a single particle 
inside the ring and solve them analytically. We find that the importance of the shear caused by variation of 
the radial velocity component with local vertical direction cannot be ignored and it may be responsible 
for damping of the bending waves in planetary rings as observed by the Voyager data. We present the wave 
profile resulting from the dissipation. We estimate that the surface mass density of the C ring to be of the 
order of $\sigma \sim  1.2-1.6$gm cm$^{-2}$, and the height $h \sim  2.2-2.4$m. These theoretical 
results are in agreement with observations.
\end{abstract}


\noindent ACCEPTED FOR PUBLICATION IN MNRAS

\section{Introduction}

Voyager photo-polarimeter (PPS) experiment indicated that the thickness of the Saturn's ring is probably 
smaller than $200m$ at the outer edge of the A -ring (Lane {\emph{et al}}, 1982). When the 
oscillatory features in bending waves are identified with possible resonance events (Shu {\emph{et al}},
1983), after a fit of the ring profile, the effective thickness is found. The effective thickness
of the ring is found to be only a few tens of meters (Rosen et al. 1988, 1991), 
Chakrabarti \cite{chak89} and Chakrabarti and Bhattacharyya \cite{chak01}. One of the important new ingredients
that has gone into estimation of the thickness is a new source of shear which was found to be present during 
detailed numerical study of dynamics of particles inside the Saturn's ring. This extra source of shear is 
found to effectively damp out the {\it{5:3}} bending wave of Mimas within a few tens of kilometers and since then it 
is considered to be important (Nicholson {\emph{et al}} 1991, Rosen {\emph{et al}}, 
1991 and Brophy {\emph{et al}}, 1992) to describe correctly the particle dynamics inside a ring. 

We now present briefly some of the work done on vertical structure and wave profile prior to ours.
Simon and Jenkins \cite{sj94} performed their analysis assuming the system having a dilute flow maintained by
identical smooth spherical frost-free particles. The balance laws were obtained
by taking moments of the Boltzmann equation. While taking care of collisions between particles,
they assumed the normal velocity components, before and after the collision, to be 
related through the coefficient of restitution, while the tangential components
were unaffected. They also assumed a symmetry about the equatorial plane ($z=0$)
and that there was no mean motion normal to the mid-plane. 
They concentrated on the outer edges of the Saturn's A -ring 
and used parameters near {\it{6:7}} Lindblad resonance to obtain relations between granular parameters
and the co-efficient of restitution and optical depth. 
They find the coefficient of restitution to be monotonically increasing with the optical depth.
No effort was made to give the profile of the bending wave.

Salo \cite{salo91} studied viscous stability properties of dense planetary rings with numerical simulations.
He confirmed earlier results of Wisdom and Tremaine \cite{wt88} that for the standard elastic model of icy particles,
the viscous instability is not expected for identical meter-sized particles, while 
for denser systems with centimeter sized particles the possibility persists. Salo also did simulations including
self-gravity resulting in the variation of particle number density with vertical height. In this simulation, Salo
considered $100$-$200$ impacts per particle. He assumed that the particle density is the highest
on the mid-plane of the ring. This simulation may be applicable for Saturn's E, F and G ring, Uranian and Jovian rings.
Later, Salo \cite{salo92} did numerical simulations of collisional systems with power law 
distribution of particle sizes and found that equilibrium geometric thickness of the Saturn's ring to 
be around $25$m for layers of cm-sized particles, and around $10$m for larger particles.

Schmidt {\emph et al.} \cite{schm99} investigated viscous oscillatory over-stability of an 
unperturbed dense planetary ring which may play a role in the 
formation of radial structure in Saturn's B ring. They extended the modeling of Simon and Jenkins \cite{sj94}
and compared Gaussian ansatzs which were used in Goldreich \& Tremaine \cite{gold78} and Araki \& Tremaine \cite{araki86}
with hydrodynamical ansatzs. They considered enhanced vertical 
frequency which appears to flatten the disk and enhances collision frequency. 
The simulation was carried out with identical sized particles. This simulation shows initial onset 
of over-stability in the ring and leads to the development of radial structure although no explanation was
presented on the damping of the bending wave.

In our paper, we are considering approximate analytical equations obtained
from the kinematic formulation for Titan {\it{-1:0}} vertical resonance to estimate a relation between
surface mass density, ring height and the damping length of the bending wave in the C -ring. Our approach
emphasizes the effect of vertical motion of particles over and above the epicyclic one. 
While doing the vertical and horizontal excursion, the particles collide and transport 
vertical component of momentum, thereby damping the vertical profile of the wave.
In the present paper, we write down the equations governing the particle dynamics including this
form of shear. We compute analytically the shear developed for Titan {\it{-1:0}} bending 
resonance and then study the damping length as a function of surface 
mass density and height as free parameters. We restrict our attention to that region of the parameter space where the 
properties of the wave profile is comparable to those observed. There are degeneracies in the parameter space 
in the sense that the same damping length could be achieved for high and low surface mass densities 
($\sigma$). Upon scrutiny it was observed that one of the solutions (that corresponding to lower $\sigma$) 
created too many wave profiles than what is observed and thus it was rejected. 
The present analysis provides the wave profile and   
therefore a comparison with the observation is easier to make. Our earlier 
(Chakrabarti and Bhattacharyya, 2001) procedure could give population distribution throughout the ring 
at different phases which is not possible in the present method. A realistic three dimensional simulation 
would probably be essential to understand the correct description of the problem.

It is to be noted that we have included only the effect of energy loss due to infrequent collision
among shearing layers and no attempt has been made to include cooling effects because of poorly
understood physics. The collisional damping reduces kinetic energy and the degree of vertical 
excursion.  This is used at each radius to obtain the wave profile. 
We computed the coefficient of restitution in our ring and found this
to be close to unity, in near agreement with the results of Simon and Jenkins \cite{sj94} 
where the coefficient is found to be $0.9$ for the same disk parameter. In other words,
we assumed nearly-elastic collision among the particles. In future, we shall model the physics 
of breakup and merger of `particles' in the ring and other associate effects. 

In the next section, we present the basic equations of the motion of particle and explain the genesis of each 
of the terms. In $\S$3, we present the solutions of the equations and compute the average shear analytically. 
In $\S$4, we apply our results in various ring conditions. Finally, in $\S$5, we draw our conclusions.

\section{Model equations}

\begin{figure}
\begin{center}
\includegraphics[height=6truecm,width=16truecm]{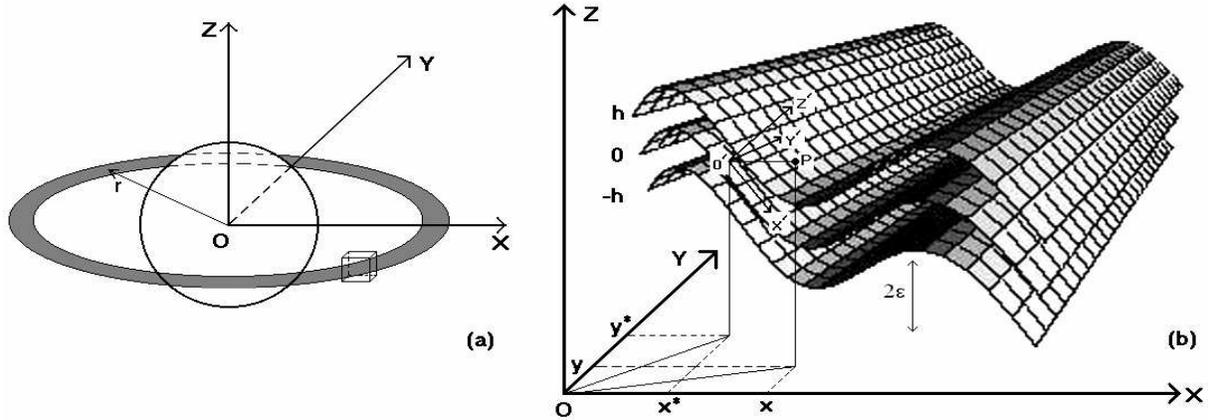}
\end{center}
\caption{Choice of the co-ordinates in our calculations. The broad view with the planet and the ring 
is shown in (a). In (b) we show details of the co-ordinates used in the small box of (a).
Here, the mid-plane of the bending wave is shown to be flanked by two artificial `boundaries' 
located at $\pm h$ distance from the mid-plane. The particle P is located at (x',y',z')
in the co-ordinate system with origin at (x$^*$, y$^*$, z$^*$). See text for details.
\label{fig:ring}
}
\end{figure}

We choose the usual right handed Cartesian coordinate system ($X$, $Y$, $Z$) at a radial distance $r$ from 
the center of the planet with the origin located on the equatorial plane of the planet with $X-Y$ plane 
coinciding with this plane (See, Fig. \ref{fig:ring}).

The $X$ -axis points radially outward and $Y$ -axis points toward the azimuthal direction. The frame is 
rotating around the planet  with the local Keplerian frequency $\Omega(r)$. Let the amplitude of the bending 
wave be $\epsilon$. The mid-plane itself oscillates up and down with this amplitude. Matter 
oscillates up and down around the instantaneous mid-plane with an amplitude $h'=h/2$ -- half
of the thickness $h$ of the ring. Let the coordinate of the origin $O^{\prime}$ of a Cartesian frame ($X^{\prime}$, 
$Y^{\prime}$, $Z^{\prime}$) which is oscillating with the mid-plane of the disk be ($x^*$, $y^*$, $z^*$). In 
the absence of oscillations of the mid-plane, these two coordinate systems merge. A particle moving within 
the ring having coordinate ($x$, $y$, $z$) has a coordinate of $x_1$=$x$-$x^*$, $y_1$=$y$-$y^*$ and 
$z_1$=$z$-$z^*$ with respect to $O^{\prime}$. If $\omega$ be the disturbance frequency due to moon forcing,
i.e., the angular frequency of the propagating wave, then the phase of the mid-plane is $\phi^*$=$k_x x^*$+$k_y y^*$-$\omega t$ 
and that of the particle located at a point $P$ ($x$, $y$, $z$) is $\phi$=$k_x x$ + $k_y y$ - $\omega t$ 
(Here $k_x$ and $k_y$ are the $x-$ and $y-$ components of the wave vector $\vec{k}$.)
so that the phase difference between the particle and the midplane is:
\begin{eqnarray}\label{eq:fifistar}
\phi - \phi^*=k_x \left( x - x^* \right) + k_y \left( y - y^* \right)=k_x x_1 + k_y y_1.
\end{eqnarray}
The frequency $\omega$ has a linear relation with orbital frequency of moon around the 
planet ($\Omega_M$), vertical ($\mu_M$) and epicyclic ($\kappa_M$) frequencies of moon given by:
\begin{equation}
\omega = m \Omega_M \pm n \mu_M \pm p \kappa_M, \label{eq:omega} .
\end{equation}
where, $m$, $n$ and $p$ are non-negative integers. For small perturbations from a satellite, the test 
particle responds as a multi-dimensional, forced, linear-harmonic
oscillator. In particular, it may suffer resonances if the relative frequency at which it experiences the
disturbance of the satellite referred to its local rotation rate, is equal to any of the natural frequencies of
its free oscillations as discussed by Franklin and Colombo \cite{frcol}. The test particle will suffer vertical 
resonance due to inclination of orbit of satellite if its radial distance of vertical resonance ($r_V$) from 
the center of the planet satisfies
\begin{equation}
\omega-m \Omega(r_V)=\pm \mu(r_V).
\end{equation}
We can always consider $\Omega_M < \Omega$ because the large satellites orbit outside the main 
ring system. Here, $m=\Omega(r)/\Omega_M$ designates a particular 
resonance (See, e.g., Shu, 1984 and Murray and Dermott, 1999 for details.). Satellites will impart forcing 
of different strength at different orbits and this is taken care of by putting different 
values of $m$, $n$ and $p$ in Eq. \ref{eq:omega}. We considered Titan {\it -1:0} resonance 
for which $m=1$ in used in Eq. \ref{eq:omega}. This mode is the strongest
resonance imparted by Titan on Saturn's ring. 
By definition, $\omega$ = ($\Omega_P$-$\Omega$), where, $\Omega_P$ is the pattern frequency. 
In the present resonance, this is the angular frequency of the perturbing moon with an opposite sign 
of orbital frequency of moon.  It is to be kept in mind that Titan has a semi-major axis of about 
$1,221,850$ km to revolve round Saturn along an orbit having an orbital inclination of $0.33$ degree with
Saturnian equatorial plane. Due to this inclination of orbit, spiral bending wave is launched in the ring at 
the resonance location $r_V=77514.8$ km which is inside Saturn's C ring. Hence the properties of the C ring 
could be obtained by understanding the behaviour of the bending wave due to Titan {\it -1:0} resonance.

Let ($x^{\prime}$, $y^{\prime}$, $z^{\prime}$) be the coordinate of the particle in the ($X^{\prime}$, $Y^{\prime}$, 
$Z^{\prime}$) frame. Thus, 
\begin{equation}
z^{\prime}=h+\epsilon \cos \phi^*,
\end{equation}
\noindent where, $h$ is the half-thickness of the ring as stated earlier. It is easy to show that the local normal 
at ($x^{\prime}$, $y^{\prime}$, $z^{\prime}$) on the mid-plane is given by, 
\begin{equation}
\hat{n}=-\hat{i} \frac{\epsilon k_x \sin \phi^*}{W}- \hat{j}\frac{\epsilon k_y \sin \phi^*}{W}-\hat{k} \frac{1}{W},
\end{equation}
where, $W$=$(1+\epsilon^2 k^2 \sin \phi^*)^{1/2}$. From this, one derives the following important relations,
\begin{eqnarray}
x_1 &=& z_1 \epsilon k_x \sin \phi^*, \\
\mbox{and }\ \ \ y_1 &=& z_1 \epsilon k_y \sin \phi^*.
\end{eqnarray}

It is conventional to write the potential $\phi_P$  due to the planet in the equatorial plane as
\begin{equation}
\phi_P(r,0)=-\frac{G M_P}{r} \left[ 1- \sum_{n=1}^{\infty} J_{2n} \left(R_P/r\right)^{2n} P_{2n}(0) \right],
\end{equation}
where, $M_P$ is mass of the planet, $R_P$ is its radius, $J_{2n}$ is its $2n^{th}$ multipole moment, $P_{2n}(0)$
is the Legendre polynomial of order $2n$ at the origin.
If one ignores the oblateness of the planet, then the multiple moment terms
are absent and the vertical frequency $\mu$ of the particle defined by,
\begin{equation}
\mu^2=\frac{\partial ^2 \phi_P}{\partial z^2}
\end{equation}
becomes identical to the Keplerian frequency. The epicyclic frequency $\kappa$ of the particle defined by,
\begin{equation}
\kappa^2=\frac{1}{r^3}\frac{d}{dr}\left[\left(r^2 \Omega \right)^2\right].
\end{equation}
This is also identical to the local Keplerian frequency $\Omega$.

In general, there will be three major acceleration terms,
\begin{equation}
\vec{g}=\vec{g}_P+\vec{g}_D+\vec{g}_M ,
\end{equation}
where, subscripts $P$, $D$ and $M$ denote the acceleration due to the planet, the self-gravity of the ring and the moon
respectively. In the first approximation, one can assume that the vertical motion is due to the planet and the moon only, 
so that the vertical component of the equation of the motion is given by,
\begin{equation}
\ddot{z}=-\Omega^2 z + a \cos \left( k_x  x + k_y y - \omega t \right).
\end{equation}
The solution of this equation is,
\begin{equation}
z=-\frac{a}{\Omega^2-\omega^2} \cos \left( k_x x + k_y y - \omega t \right). \label{eq:zsol}
\end{equation}
The factor in front of the cosine term can be identified with the amplitude $\epsilon$ of the vertical movement of 
the mid-plane due to the satellite forcing provided $\epsilon \left(\Omega^2 - \omega^2 \right)=a$. 
Thus the forcing term due to the satellite is,
\begin{equation}
g_M=\epsilon \left(\Omega^2-\omega^2\right) \cos \left( k_x x + k_y y - \omega t \right).
\end{equation}
Its components are to be added in the differential equation governing the motion of the particle.
We are not interested in the solution of the homogeneous equation (Eq. 12) since it would be periodic and 
would average out.

Before proceeding further, let us try to derive the transformation rule for vectors between the reference frames.
In the warped ring having the frame of reference $X^{\prime}Y^{\prime}Z^{\prime}$, 
\begin{eqnarray}\label{eq:zp}
z' &=& h+\epsilon\rm{cos} \phi^*, \nonumber \\
   &=& h+\epsilon \rm{cos}\left(k_xx'+k_yy'-\omega t \right),
\end{eqnarray}
From this we derive,
\begin{eqnarray}
                & & \frac{\partial z'}{\partial x'} = -\epsilon k_x \sin \phi^*, \\
                & & \frac{\partial z'}{\partial y'} = -\epsilon k_y \sin \phi^*, \\
		& & \frac{\partial z'}{\partial z'} = 1.
\end{eqnarray}

A unit vector normal to surface described by Eq. \ref{eq:zp} is written as,
\begin{eqnarray}
\hat{k'} &=& \frac{1}{\sqrt{\left(\partial z'/\partial x'\right)^2+\left(\partial z'/\partial y'\right)^2
+\left(\partial z'/\partial z'\right)^2}} \ \left[\hat{i}\frac{\partial z'}{\partial x'}+
\hat{j}\frac{\partial z'}{\partial y'}+\hat{k}\frac{\partial z'}{\partial z'}\right], \nonumber
\end{eqnarray}
\begin{eqnarray}\label{eq:kp}
\mbox{or, }&&\hat{k'}= \hat{i}\left[-\frac{\epsilon k_x \sin \phi^*}{\sqrt{1+\epsilon^2k^2\sin^2\phi^*}}\right]
+\hat{j} \left[- \frac{\epsilon k_y\sin\phi^*}{\sqrt{1+\epsilon^2k^2\sin^2\phi^*}}\right] \nonumber \\
&+&\hat{k}\left[\frac{1}{\sqrt{1+\epsilon^2k^2\sin^2\phi^*}}\right] .
\end{eqnarray}
From Fig. \ref{fig:ring}, 
\begin{eqnarray}
\vec{O'P} &=& (x-x^*)\hat{i}+(y-y^*)\hat{j}+(z-z^*)\hat{k} \nonumber \\
&=& \hat{i} x_1+\hat{j} y_1+\hat{k} z_1, \\
\mbox{and } \nonumber \\
\left|O'P \right| &=& \sqrt{x_1^2+y_1^2+z_1^2}.
\end{eqnarray}
The unit vector is given by,
\begin{eqnarray}
\frac{\vec{O'P}}{|\vec{O'P}|} &=& \hat{i} \frac{x_1}{\sqrt{x_1^2+y_1^2+z_1^2}}
+\hat{j}\frac{y_1}{\sqrt{x_1^2+y_1^2+z_1^2}}+\hat{k}\frac{z_1}{\sqrt{x_1^2+y_1^2+z_1^2}}. \nonumber \\
&=& \vec{k'}.
\end{eqnarray}

These relations yield,
\begin{eqnarray}\label{eq:kpcomp1}
\frac{x_1}{\sqrt{x_1^2+y_1^2+z_1^2}} = - \frac{\epsilon k_x \sin \phi^*}
{\sqrt{1+\epsilon^2k^2\sin^2\phi^*}},
\end{eqnarray}

\begin{eqnarray}\label{eq:kpcomp2}
\frac{y_1}{\sqrt{x_1^2+y_1^2+z_1^2}} = - \frac{\epsilon k_y \sin \phi^*}
{\sqrt{1+\epsilon^2k^2\sin^2\phi^*}},
\end{eqnarray}

\begin{eqnarray}\label{eq:kpcomp3}
\frac{z_1}{\sqrt{x_1^2+y_1^2+z_1^2}} = \frac{1}{\sqrt{1+\epsilon^2k^2\sin^2\phi^*}}.
\end{eqnarray}

From eqns. (\ref{eq:kpcomp1})  and  (\ref{eq:kpcomp3}) we obtain,
\begin{eqnarray}
\frac{x_1}{z_1}=-\epsilon k_x\sin \phi^*, \nonumber \\
or, \   x_1=z_1 \epsilon k_x \sin \phi^*.  \label{eq:x1}
\end{eqnarray}

Similarly, from (\ref{eq:kpcomp2}) and (\ref{eq:kpcomp3}),
\begin{eqnarray}
\frac{y_1}{z_1}=-\epsilon k_y \sin \phi^*, \nonumber \\
or,\  y_1=z_1 \epsilon k_y \sin \phi^*.   \label{eq:y1}
\end{eqnarray}

From eqns. (\ref{eq:fifistar}, \ref{eq:x1},  \ref{eq:y1}),
\begin{eqnarray}\label{eq:fifistar1}
\phi - \phi^* &=& k_x \left( z_1 \epsilon k_x \sin \phi^* \right) + k_y \left( z_1 \epsilon k_y \sin \phi^*
\right) \nonumber \\
&=& z_1 \epsilon \left( k_x^2 \sin \phi^* + k_y^2 \sin \phi^* \right) \nonumber \\
&=& z_1 \epsilon k^2 \sin \phi^*.
\end{eqnarray}

Following techniques used by Shu \cite{shu84}, the amplitude in the radial direction can be assumed
to have the form, $a(x)=a_0 exp \left[ i \int k_x dx \right]$. 
To get the principal Fourier component of the undamped disk, we choose
$a(x)=a_0 exp \left[ ik_x x \right]$. Identifying the wave amplitude with $\epsilon$, one can clearly write $a_0$
as $a_0= \epsilon \left(\Omega^2- \omega^2 \right)$.

Therefore, from eqns.(\ref{eq:zsol}), the force component for satellite forcing is given as,
\begin {eqnarray}\label{eq:znokp}
z_{\rm{sat}}=\epsilon \left(\Omega^2-\omega^2 \right)\cos(k_x x+k_y y -\omega t).
\end{eqnarray}

To get the components of satellite forcing from eqn. (\ref{eq:znokp}) in the frame of reference placed at the mid-plane
of warped ring system, one has to use the properties given in eqn.(\ref{eq:kpcomp1}-\ref{eq:kpcomp3}). 
The components of the equation of motions of the test particle are then given by,
\begin{eqnarray}
\frac{d^2x}{dt^2} &=& -2 \Omega \frac{dy}{dt}+3 \Omega^2 x -\nu^2 x_1 -\left( \Omega^2-\omega^2 \right) 
\epsilon^2 k_x  \cos \phi \sin \phi^*/W, \label{eq:xeq}\\
\frac{d^2y}{dt^2} &=& 2 \Omega \frac{dx}{dt} -\nu^2 y_1 -\left( \Omega^2-\omega^2 \right) \epsilon^2 k_y 
\cos \phi \sin \phi^*/W, \label{eq:yeq}\\
\frac{d^2z}{dt^2} &=& -\Omega^2z -\nu^2 z_1+\left( \Omega^2-\omega^2 \right) \epsilon \cos \phi / W. \label{eq:zeq}
\end{eqnarray}
Here, $\nu^2=4 \pi G \rho$ is the vertical oscillation frequency due to the self-gravity and $\rho$ is 
the mass density of the ring matter. The first terms in the R.H.S. of Eq. (30) and (31) are the Coriolis 
acceleration term. The second term of eq. (30) comes from the difference between the centrifugal 
acceleration of the particle and the centrifugal acceleration of the observer in rotating frame of 
reference. The vertical forcing is associated with the coordinates in the primed (warped) frame.
We consider the case when $\epsilon \kappa \ll 1$. i.e., for wavelengths 
which are large compared to the amplitude of the bending wave. In this limit, 
$W \approx 1$ and $\phi =\phi^*+z_1 \epsilon \kappa^2 \sin 
\phi^*$.  This means $\cos \phi \approx \cos \phi^*-z_1 \epsilon \kappa^2 \sin^2 \phi^*$. 

In Eqs. (\ref{eq:xeq}-\ref{eq:zeq}), we considered a kinematic approach, and not 
the fluid dynamical approach. That is, we are not using Navier-Stokes' equation. 
This is because in the C ring which is made up of the neutrals,
particles collide with one another very scarcely, once or twice in each orbital revolution  
(This many not be true in the partially ionized rings like E, F and G, where charged particles coupled to the magnetic 
field gyrate and the gyro-radius is much smaller than the ring thickness reducing the mean free-time
and mean free-path drastically). Thus we prefer to treat the  C ring as a collisionless system 
with particles satisfying Boltzmann distribution. However, once we obtain the local shear, we can use this
to compute the local dissipation when the mean collision time is supplied. This two-step process has been adopted
in our study.

We first re-rite the $z$-component of the equation as,
\begin{equation}
\frac{d^2 z_1}{dt^2}=-\alpha^2 z_1-\epsilon^2 \left( \Omega^2-\omega^2 \right) z_1 k^2 \sin^2 \phi^*,
\end{equation}
where, $\alpha^2=\Omega^2+\nu^2$. It is easy to transform the equation in terms of the phase $\phi^*$ by 
using $\partial/\partial t \equiv  - \omega \partial/\partial \phi^*$ and $\partial^2/\partial t^2 \equiv 
\omega^2 \partial^2/ \partial \phi^{*2}$. This yields,
\begin{equation}\label{eq:eqz1a}
\frac{\partial^2 z_1}{\partial \phi^{*2}} =- \frac{z_1}{\omega^2} \left[ \alpha^2 + \frac{\epsilon^2 
\left( \Omega^2- \omega^2 \right) k^2}{2}\right]+\frac{\epsilon^2 \left( \Omega^2 - \omega^2 \right) z_1 k^2 
\cos 2 \phi^*}{2 \omega^2}.
\end{equation}
Close to the resonance orbit, $\Omega \sim \omega$ and the parameter,
\begin{equation}
\gamma^2 = \frac{\epsilon^2 k^2 \left( \Omega^2 - \omega^2 \right)}{\omega^2}
\end{equation}
could be treated as a perturbation parameter (Chakrabarti, \cite{chak89}). Defining,
\begin{equation}
\eta^2=\frac{\alpha^2}{\omega^2}+\frac{\gamma^2}{2},
\end{equation}
Eq. (\ref{eq:eqz1a}) becomes,
\begin{equation}\label{eq:eqz1b}
\frac{\partial ^2 z_1}{\partial \phi^{*2}}=-\eta^2 z_1+\frac{\gamma^2 \cos 2 \phi^*}{2}z_1.
\end{equation}
To obtain a complete solution, we expand $z_1$ in powers of $\gamma^2$ as,
\begin{equation}
z_1=z_1^{(0)}+\sum_{i=0}^N \gamma^{2i} z_1^{(i)}.
\end{equation}
Upon substitution in Eq. \ref{eq:eqz1b} and equating coefficients of the powers of $\gamma^2$ we obtain, up to
powers of $\gamma^4$, the solution of $z$ as,
\begin{eqnarray}\label{eq:eqzfinal}
z&=&\epsilon \cos \phi^* +h \sin \eta \phi^* -\frac{h \gamma^2}{16} \left[ \frac{\sin \left( 2+\eta 
\right)\phi^*}{1+\eta}-\frac{\sin\left(2-\eta\right)\phi^*}{1-\eta}\right] + \frac{h \gamma^4}{64} \left[ 
\frac{\sin \left( 4+\eta \right) \phi^*}{8\left(1+\eta\right)\left(2+\eta\right)} \right.\nonumber \\
& &\left.-\frac{\sin \eta \phi^*}{\left( 1+ \eta \right)\left(1-\eta \right)}
-\frac{\sin\left(4-\eta\right)\phi^*}{8\left(1-\eta\right)\left(2-\eta\right)}\right].
\end{eqnarray}
From Eqs. \ref{eq:zeq} and \ref{eq:yeq}, we can now obtain the solutions for $x$ and $y$,
\begin{eqnarray}\label{eq:xfinal}
x&=&h \sin \left( \frac{\Omega \phi^*}{\omega} \right) + \frac{\nu^2 \epsilon k_x h}{2 \omega^2} 
\left[\frac{1+\gamma^2/(16(1-\eta))}{\Omega^2/\omega^2-(1-\eta)^2} \cos(1-\eta)\phi^* \right. \nonumber \\
& &-\frac{1+\gamma^2/(16(1+\eta))}{\Omega^2/\omega^2-(1+\eta)^2} \cos(1+\eta)\phi^* 
+\frac{\gamma^2/16+\gamma^4/(512(2-\eta))}{(1-\eta)[\Omega^2/\omega^2-(3-\eta)^2]} \cos(3-\eta)\phi^* \nonumber \\
& &+\frac{\gamma^2/16+\gamma^4/(512(2+\eta))}{(1+\eta)[\Omega^2/\omega^2-(3+\eta)^2]} \cos(3+\eta)\phi^* 
+\frac{\gamma^4/(512(1-\eta)(2-\eta))}{[\Omega^2/\omega^2-(5-\eta)^2]} \cos(5-\eta)\phi^* \nonumber \\
& &\left.-\frac{\gamma^4/(512(1+\eta)(2+\eta))}{[\Omega^2/\omega^2-(5+\eta)^2]} \cos(5+\eta) \phi^* \right]
\end{eqnarray}
and
\begin{eqnarray}\label{eq:yfinal}
y &=& 2 h \cos \left( \frac{\Omega \phi^*}{\omega} \right) - \frac{\nu^2 \epsilon k_x h \Omega}{\omega^3}
\left[ \frac{1+\gamma^2/16(1-\eta)}{\Omega^2/\omega^2-(1-\eta)^2} \frac{1}{(1-\eta)} \sin (1-\eta) 
\phi^* \right. \nonumber\\
& & +\frac{1+\gamma^2/16(1+\eta)}{\Omega^2/\omega^2-(1+\eta)^2}\frac{1}{(1+\eta)}\sin (1+\eta) \phi^* \nonumber \\
& & +\frac{\gamma^2/16+\gamma^4/512(2-\eta)}{\Omega^2/\omega^2-(3-\eta)^2}\frac{1}{(1-\eta)(3-\eta)}
\sin(3-\eta)\phi^*\nonumber\\
& & +\frac{\gamma^2/16+\gamma^4/512(2+\eta)}{\Omega^2/\omega^2-(3+\eta)^2}\frac{1}{(1+\eta)(3+\eta)}
\sin(3+\eta)\phi^*\nonumber\\
& & +\frac{\gamma^4/512(1-\eta)(2-\eta)}{\Omega^2/\omega^2-(5-\eta)^2}\frac{1}{(5-\eta)(3-\eta)}
\sin(5-\eta)\phi^*\nonumber\\
& & -\left.\frac{\gamma^4/512(1+\eta)(2+\eta)}{\Omega^2/\omega^2-(5+\eta)^2}\frac{1}{(5+\eta)(3-\eta)}
\sin(5+\eta)\phi^*\right]\nonumber\\
& & +\frac{\nu^2\epsilon k_y h}{2 \omega^2}\left[\frac{\cos(1-\eta)\phi^*}{(1-\eta)^2}-
\frac{\cos(1+\eta)\phi^*}{(1+\eta)^2}-\frac{\gamma^2\cos(1+\eta)\phi^*}{16(1+\eta)^3} \right. \nonumber \\
& & \left.+\frac{\gamma^2 \cos(3+\eta)\phi^*}{16(1+\eta)(3+\eta)^2}+ \cdots 
\mbox{other terms of higher order} \right].
\end{eqnarray}
Numerical simulations shown by Chakrabarti \cite{chak89} and Chakrabarti \& Bhattacharyya \cite{chak01} tend 
to indicate that the shear due to variation of radial velocity component along vertical direction is 
significant. However, for the sake of completeness, the variation of vertical velocity component along the 
radial direction is also considered. Analytically, the magnitude of this shear $S$ could be computed from,
\begin{equation}
S^2 \sim \left( \frac{<\ddot{x}>}{<\dot{z}>}\right)^2 + \left( \frac{<\ddot{z}>}{<\dot{x}>}\right)^2,
\end{equation}
where, we used only the significant components in a corotating frame and $<>$ indicates that averaging is being 
done over a complete cycle $\phi^*$. To show how the shear looks like analytically, we show here  the nature of the
first term of Eq. (23),

\begin{equation}
\frac{<\ddot{x}>}{<\dot{z}>}= \frac{P}{Q},
\end{equation}

where, \\
\begin{eqnarray}
P&=&h\Omega \omega \left[ \cos \left(\frac{2 \pi \Omega}{\omega}\right)-1\right] \nonumber \\
&-&\frac{\nu^2\epsilon k_x h}{2}\left[\left\{\frac{1+\gamma^2/16(1-\eta)}{\frac{\Omega^2}{\omega^2}-(1-\eta)^2}(1-\eta)^2
+\frac{1+\gamma^2/16(1+\eta)}{\frac{\Omega^2}{\omega^2}-(1+\eta)^2}(1+\eta)^2\right\}\right. \nonumber \\
& &\frac{(-\eta \sin 2\pi \eta)}{2 \pi(1+\eta)(1-\eta)}+\left\{\frac{1+\gamma^2/16(1-\eta)}{\frac{\Omega^2}{\omega^2}-(1-\eta)^2}
(1-\eta)^2-\right. \nonumber \\
&&\left.\frac{1+\gamma^2/16(1+\eta)}{\frac{\Omega^2}{\omega^2}-(1+\eta)^2}(1+\eta)^2\right\}\frac{(-\sin 2\pi\eta)}{2\pi(1+\eta)(1-\eta)}
\nonumber \\
&+&\left\{\frac{\gamma^2/16+\gamma^4/512(2-\eta)}{(1-\eta)\left\{\frac{\Omega^2}{\omega^2}-(3-\eta)^2\right\}}
(3-\eta)^2\right.\nonumber \\
&+&\left.\frac{\gamma^2/16+\gamma^4/512(2+\eta)}{(1+\eta)\left\{\frac{\Omega^2}{\omega^2}-(3+\eta)^2\right\}}(3+\eta)^2\right\}
\frac{(-\eta \sin 2\pi \eta)}{2\pi(3+\eta)(3-\eta)} \nonumber \\
&&+\left\{\frac{\gamma^2/16+\gamma^4/512(2-\eta)}{(1-\eta)\left[\frac{\Omega^2}{\omega^2}-
(3-\eta)^2\right]}(3-\eta)^2 \right.\nonumber \\
&&\left.-\frac{\gamma^2/16+\gamma^4/512(2+\eta)}{(1+\eta)\left[\frac{\Omega^2}{\omega^2}-(3+\eta)^2\right]}(3+\eta)^2\right\}
\frac{(-\sin 2 \pi \eta)}{2 \pi(3+\eta)(3-\eta)}\nonumber \\
&&+\frac{\gamma^4/512(1-\eta)(2-\eta)}{\frac{\Omega^2}{\omega^2}-(5-\eta)^2}(5-\eta)^2\left\{\frac{-\eta \sin 2 \pi \eta}
{2 \pi (5+\eta)(5-\eta)}\right.\nonumber \\
&& \left.\left.- \frac{\sin 2\pi \eta}{2 \pi (5+\eta)(5-\eta)}\right\}\right], \nonumber \\
\end{eqnarray}
and, \\
\begin{eqnarray}
Q &=& \frac{-h\omega}{2 \pi}\sin 2\pi\eta\nonumber \\
&-&\frac{h\gamma^2\omega}{8}\left[\frac{-\eta^2 \sin 2 \pi \eta}{2\pi(1-\eta^2)(4-\eta^2)}+\left(\frac{2-\eta^2}{1-\eta^2}\right)\right]
\nonumber \\
&-&\frac{-h\gamma^4\omega}{64}\left[\frac{(\eta^2+8)\sin 2\pi \eta}{8(1-\eta^2)(4-\eta^2)(16-\eta^2)} \right.\nonumber \\
&-&\left.\frac{\eta^2(\eta^2-10)\sin2\pi\eta}{8\pi(1-\eta^2)(4-\eta^2)(16-\eta^2)}-\frac{\eta\sin 2 \pi \eta}
{2\pi(1-\eta^2)} \right] . \nonumber  \\
\end{eqnarray}

During the excursion of the particle in epicyclic orbits as well as in vertical motion inside the ring, the particle
is expected to collide with a neighboring particle at least once in each orbit and in this process
the vertical component of momentum is transferred. In this 
context we wish to remind that it is not enough just to have a differential motion, the collision must also takes
place in order that the momentum is transferred. Shu \cite{shu84} already pointed out that 
an anomalous viscosity may be required if one has to use only the differential motion due to the Keplerian
distribution. The contribution $\psi=-\frac{3}{2}\Omega$ to the $r-\phi$ component of the
shear stress $S_{r\phi}=\eta \psi = \eta r \frac{\partial \Omega}{\partial r}$, where $\eta$ is the 
co-efficient of viscosity. This contribution at the resonance radius that we are considering is
$\psi=4.2\times 10^{-4}$ comparable to the shear we computed above using Eq. (23). 
However, not being a
strongly collisional system, $\eta$ need not be isotropic, and thus is likely to be larger for vertical excursion 
(due to collision among disk particles inside the ring of few meters in height) than radial
excursion. For an isotropically collisional system, the excursion in all directions
is likely to be statistically identical. In the $X-Y$ plane, such excursion due to the local temperature 
causes the epicyclic motion. For simplicity,
we assume that the excursions in the vertical and horizontal directions
are similar and the epicyclic radius is of the order of the half-height of the ring.
Thus, the mean free distance in both the vertical and the horizontal
distance is either the epicyclic radius or the size of the particle, whichever is bigger. 
In case of E and  F rings (which are not considered here), the situation may change, and  the
Keplerian  contribution could be more compared with the contribution from the vertical excursion. 

\begin{figure}
\begin{center}
\includegraphics[height=12truecm,width=12truecm]{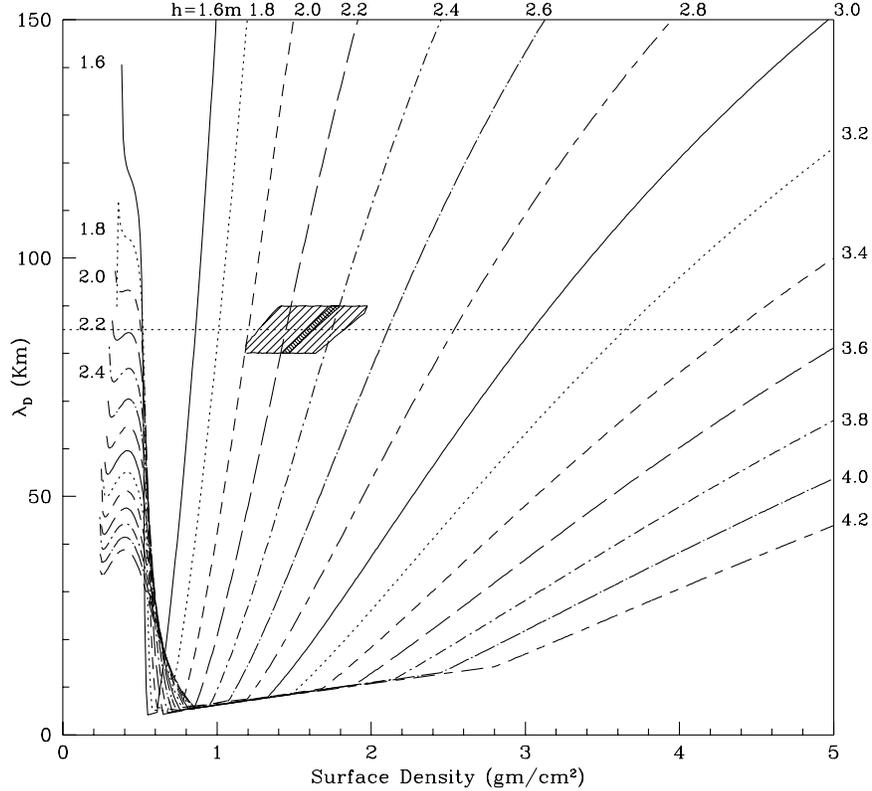}
\end{center}
\caption{Variation of the damping length ($\lambda_D$) as a function of the 
surface mass density for various ring heights ($h$). A horizontal line at $\lambda_D=85$km 
is drawn for reference. Note that there is no solution at high surface density. The shaded box indicates
a prescribed parameter zone  for the C ring as obtained from our model.
\label{fig:fig2}
}
\end{figure}

\begin{figure}
\begin{center}
\includegraphics[height=12truecm,width=12truecm]{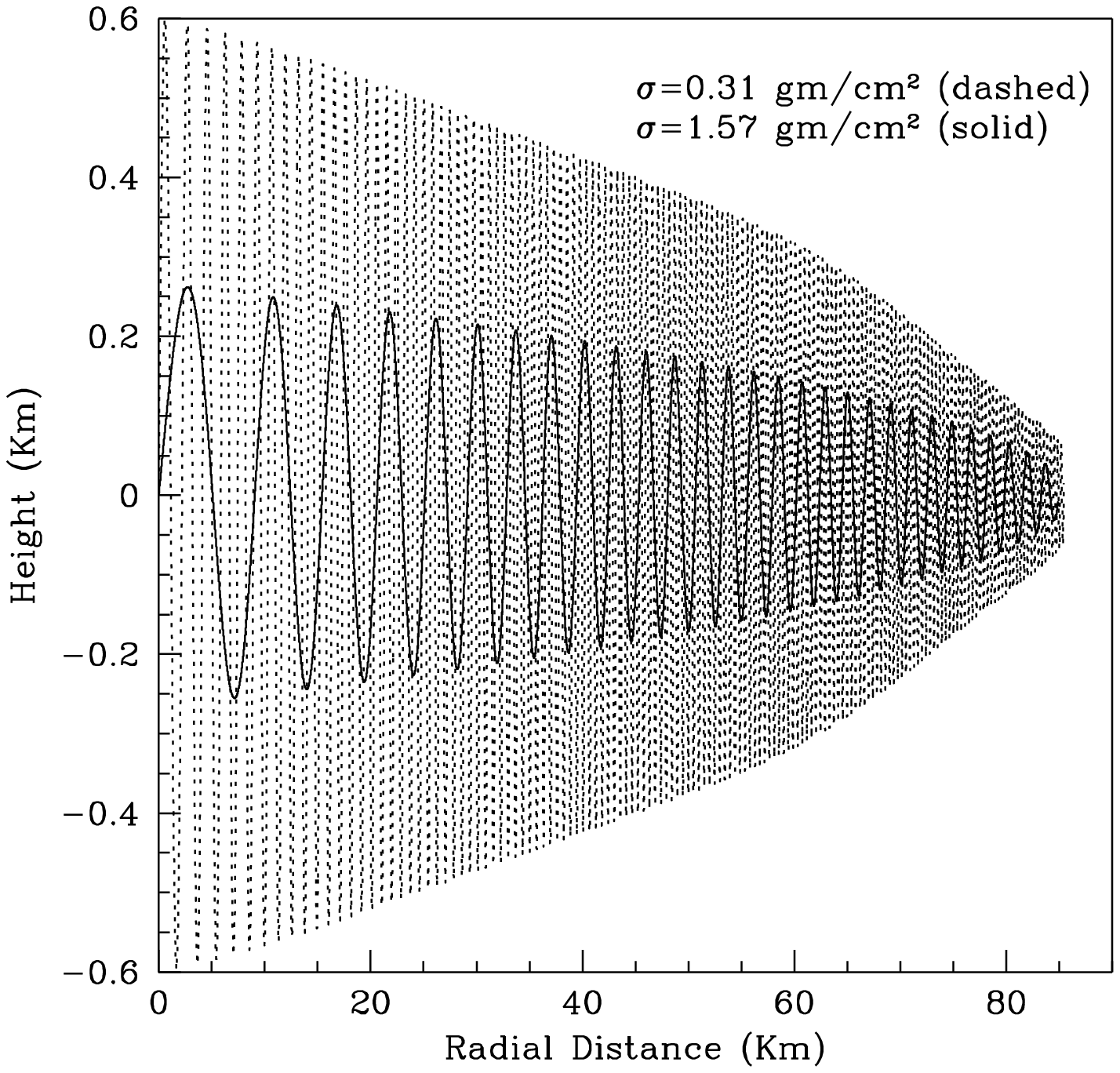}
\end{center}
\caption{The wave profiles are shown. The outer one (dashed) shows for parameters selected 
outside our prescribed shaded parameter zone as given in Fig. \ref{fig:fig2} and the number of complete waves
is few hundred while the inner one (solid) showing distinct wave-profile is obtained from a set of parameters 
taken from the shaded parameter zone. Here, the number of complete waves is $28$ as observed.
\label{fig:fig3}
}
\end{figure}

Our analytical solution may be used for the calculation of the wave profile as well as the damping length of the 
bending wave once we model the collisional process properly. Simon and Jenkins (1994), in the context of 
$A$ ring obtained the relation between the coefficient of restitution and the optical depth.
For the time being we assume that the collision is nearly-elastic, and later we shall show {\it a posteriori}
that this assumption may be justified. For a wave profile, we first compute the amplitude of the wave at the
location of the vertical resonance from Rosen {\emph{et al}} \cite{rosen91},
\begin{equation}\label{eq:epsarv}
\epsilon \left( r_v \right) = \frac{334}{\sigma^{1/2}}\ m,
\end{equation}
\noindent where, the surface density $\sigma$ is to be computed in c.g.s. units. Here $r_v$ is the location of the vertical
resonance. The radial wavelength of the launched wave is computed from Rosen {\emph{et al}}\cite{rosen91},
\begin{equation}\label{eq:lamx}
\lambda_x \left(r \right) = \frac{209 \sigma}{\left( r - r_v \right)}\ km.
\end{equation}
To calculate the damping length, we use the following procedure: we advance the wave by $\delta r$, and compute the
reduced energy density due to collisional transport at $r_2=r_1+\delta r$ from,
\begin{equation}
E_w\left(r_2\right)=E_w\left(r_1\right)-\frac{\dot{E}_w\left(r_1\right)}{c_g\left(r_1\right)} \delta r,
\end{equation}
\noindent where, the energy density of the wave is calculated using,
\begin{equation}
E_w\left(r\right)=\frac{1}{2}\epsilon^2(r)\Omega^2(r)
\end{equation}
and the rate of dissipation of the wave is obtained from,
\begin{equation}
\dot{E}_w\left(r_1\right)=\nu_k\left(r_1\right) <S>^2.
\end{equation}
The instantaneous kinematic viscosity $\nu_k(r)$ is given by,
\begin{equation}
\nu_k(r)=\Omega(r) \tau \max \left( R^2, R_{epi}^2 \right).
\end{equation}
Here, $R$ is the size of the largest particle and $R_{epi}$ is the amplitude of the epicyclic motion which is similar
to the thickness of the ring i.e., $h$. The group velocity $c_g(r)$ is given by Shu {\emph{et al}} \cite{shu},
\begin{equation}
c_g(r)=-\frac{\pi G \sigma}{\omega(r)-m\Omega(r)},
\end{equation}
where, $\omega(r)=\Omega_P-\Omega(r)$, $\Omega_P=-2\pi/T_{titan}$ is the pattern frequency, same as the angular
velocity of Titan. For {\it{-1:0}} resonance, $m=1$ was chosen.

We repeat the above procedure, till the energy of the wave vanishes. The traversed length gives 
the damping length of the bending wave. Those parameters which yielded results similar to the observed damping length 
and wavelengths may be correspond to the actual parameters of the disk.

\section{Results for various ring conditions}

In this section, we report some results of our calculation for rings of different characteristics of the Titan
{\it{-1:0}} resonance location. Computation for other rings will be studied elsewhere. The free parameters we considered are
the ring height $h$ and the surface mass density $\sigma$. In Fig. \ref{fig:fig2}, we show variation of damping 
length $\lambda_D$ (km) as a function of the surface mass density $\sigma$ for various thickness ($h$, marked
on each curve). The horizontal line is drawn to mark the observed damping length of $85$ km (Rosen et al. 1988).
This line clearly shows that a particular damping length may be observed for two different values of
surface mass density. It also shows that the ring height cannot be more than $3.6$m. The shaded region around
$\sigma \sim1.5\;$ gm cm$^{-2}$ includes those parameters for which the number of oscillations lie between $25$ and $35$
and the damping length between $80$ and $90$ km.

In order to show that the higher of the two values of $\sigma$ is really the solution, in Fig. \ref{fig:fig3} we compare
the wave profiles for the two surface mass densities such as $\sigma=0.31\; $gm cm$^{-2}$ and $1.57\;$gm cm$^{-2}$ for
ring-height of $2.3$m for which the damping length is $85$ km. It is clearly observed that the case of $\sigma=0.31\;$
gm cm$^{-2}$ (dashed curve) is invalid because the total number of oscillations is $148$, far too high compared to the
observed number. On the other hand, the profile for $\sigma=1.57\;$gm cm$^{-2}$ has 28 waves which is equal to the
observed number of oscillations as given in Rosen {\emph{et al}} \cite{rosen88}.

So far, we obtained the results assuming the collision to be nearly-elastic. In reality the co-efficient of restitution
$\epsilon_c$ need not be unity. Following simple kinematics, we can obtain the 
value of $\epsilon_c$ in terms of ring parameters from:
\begin{equation}
\epsilon_c^2=1-\frac{\nu_k <S>^2 \pi^2}{h^2 \Omega^4 \tau^2} .
\end{equation}
After putting our obtained values of $\nu_k \sim 1.2\times 10^{-5}$, $h\sim230$ cm, $\tau \sim \sigma \sim 1.57$
$gm/cm^2$, $\Omega \sim 2.835 \times 10^{-4}$, $<S> \sim 0.148 \times 10^{-4}$ we obtain
$\epsilon_c \sim 0.999297$ which is very close to unity. It is to be noted that
Simon and Jenkins (1994), using 6:7 Lindblad resonance parameters obtained the
$\epsilon_c=0.9$  for an optical depth of $1.57$ (See, Fig. 4 of that paper). 
Though there is some discrepancy with our result, what is important is that the collision is 
nearly-elastic and thus the ring parameters derived by us may be realistic.

\section{Conclusion}

Rosen {\emph{et al}} \cite{rosen88} inferred from their numerical modeling that the persistence of the wave for 
over 28 cycles indicated that the scale height of the C -ring could not be more than $2$ to $3$m. Even the radio
occultation fitting indicated it would be rather $\leq$ 2.6m for damping length of the order of $\sim85$ km. Our 
analytical calculation based on the shear component in the vertical direction shows a good agreement with the
observed results. For a given height and damping length we observed that there were multiple solutions for the 
surface density $\sigma$. We were able to eliminate one of the solutions by explicit computation of the wave profile. The 
solution with the lower surface density has far too many number of oscillations that what was observed. We find 
that the most probable value lies somewhere around $\sigma \sim 1.57\;$gm cm$^{-2}$ and height $h \sim2.3$m. We 
therefore believe that the new component of shear which was first introduced by Chakrabarti \cite{chak89} for 
Mimas $5:3$ resonance, could be operating in Titan $-1:0$ resonance as well. In future, we plan to explore other 
rings and other resonances.

It is to be cautioned that our results depends on the nature of the amplitude and wavelength at the launching site
as predicted using thin ring approximation (equations \ref{eq:epsarv}-\ref{eq:lamx}). It is possible that when the
effects of finite thickness is incorporated the parameters may vary slightly. Another limitation of our work is that 
we did not explicitly include dissipation in the equation of motion themselves to obtain the wave profiles. We solved the 
problem in a two-step manner: we first compute the shear assuming no dissipation, and then use that shear to compute dissipation 
{\it a posteriori}. Similarly, no effect due to finite size of the particles, such as 
collisional break-up, particle spin etc. were included. We plan to follow some of the works
 presented in the Introduction to incorporate these effects to remove these short-comings in future.

\section*{Acknowledgments}

AB thanks Dept. of Science and Technology for the Fast Track Young Scientist Award which enabled him to 
complete this project.

{}
\label{lastpage}
\end{document}